-----------------------------Instructions----------------------------

'

there),
file.

\documentstyle[12pt]{article}
\def\prd#1 { Phys. Rev. {\bf D#1 }} 

\def\npb#1 { Nucl. Phys. {\bf B#1 }} 

\def\plb#1 { Phys. Lett. {\bf B#1 }} 

\def\zpc#1 { Z. Phys. {\bf C#1 }} 

\def\prl#1 { Phys. Rev. Lett. {\bf#1 }}

\def\ds{ D^{(\ast)} }
\def\ra {\rightarrow }

\def\gev{ {\rm GeV} }

\begin{document}

\setlength{\baselineskip}{3.0ex}
\vspace*{2.3cm}

\hfill {Alberta Thy-39-93}
\vskip 1 cm

\begin{center}
{\large\bf 

Chiral symmetry and many-body nonleptonic decays\\
of bottom hadrons\footnote{Talk presented at the
{\it 5th International Symposium on Heavy
Flavour Physics}, Montreal, Canada, July 6-10, 1993.}
}\\
\vspace*{6.0ex}
{\large Qiping Xu\footnote{Address after Sept. 1, 1993:
Department of Physics,
University of Toronto, 60 St. George Street, Toronto, Ontario,
Canada M5S 1A7}
and A. N. Kamal} \\
\vspace*{1.5ex}
{\large\it Theoretical Physics Institute and Physics Department} \\
{\large\it University of Alberta}\\
{\large\it Edmonton, Alberta, Canada, T6G 2J1}\\
\end{center}
\vspace*{4.5ex}
\vfill
\begin{center}
{\bf Abstract}
\end{center}
\vspace*{3.0ex}
\noindent 

Chiral symmetry can be applied to many-body
nonleptonic decays of heavy hadrons.
We establish the chiral effective Hamiltonian for some
typical many-body nonleptonic decays of bottom hadrons. We
discuss the lowest-order contributions coming from such a
Hamiltonian and present, as an example, a calculation of decay 

distributions of $B^-\rightarrow D^+ D^-_s \pi^-$ and
$B^-\rightarrow D^{\ast +} D^-_s \pi^-$.
We emphasize that wide applications of this
method are possible.
\vspace*{6.0ex}
\newpage

\begin{center}
{\bf 1. Introduction}
\end{center}

Recently, there has been progress in establishing a theory
incorporating both heavy quark symmetry and chiral symmetry
to study the interactions of heavy hadrons
with Goldstone bosons \cite{r1,r2,r3}.
This method has been applied to various processes,
in particular, to semileptonic decays such as
$B \rightarrow \pi e \bar \nu$,
$B \rightarrow D\pi e \bar \nu$ \cite{r1,r2,r3}.
Though the Goldstone bosons in these decays are
quite energetic in most kinematic region, there is
always a corner on the Dalitz-plot
where the Goldstone bosons are soft and where it is
possible to apply chiral symmetry. A similar situation
also arises in many-body nonleptonic decays
of heavy hadrons involving Goldstone bosons.
Examples among these decays are, $B \rightarrow D D_s \pi$,
$\Lambda_b \rightarrow \Lambda_c D K$ etc.
So far little attention has been paid to the applications
of chiral symmetry to this class of decays.

We have applied chiral symmetry to
many-body nonleptonic decays of bottom hadrons \cite{rxk}.
Here we will give a brief illustration of our method. 

A calculation of decay distributions of
$B^-\rightarrow D^+ D^-_s \pi^-$ and
$B^-\rightarrow D^{\ast +} D^-_s \pi^-$ will be presented as
examples.

\begin{center}
{\bf 2. Chiral Effective Lagrangian for Heavy Hadrons}
\end{center}

The lowest-order chiral effective Lagrangian of heavy hadrons
(meson and baryon) which incorporate both heavy quark symmetry
and chiral symmetry is \cite{r1,r2,r3}
\begin{eqnarray}
{\cal L}^{(0)}=\!\!\!
&&-i Tr( {\bar H}_i v\cdot D H_i)+
g_1 Tr( {\bar H}_i H_j {\not\!\!A }_{ji}\gamma_5 )\nonumber\\
&&-i\ Tr_G( {\bar B^\mu_6}\ v\cdot D B_{6\mu})
-\frac{1}{2}i\ Tr_G( {\bar B_{\bar 3}}\ v\cdot D B_{\bar 3})
\nonumber\\
&&+i\ g_2 \epsilon_{\mu\nu\alpha\beta}
Tr_G ({\bar B^{T\mu}_6} v^\nu A^\alpha B^\beta_6)+
g_3 Tr_G ({\bar B_{\bar 3}} A_\mu B^\mu_6+
{\bar B^\mu_6} A_\mu B_{\bar 3} )
\label{eq:e1}
\end{eqnarray}
Here we work only in the leading order. 

The notation in (\ref{eq:e1}) is the following:
For meson field $H_i$ with heavy quark $Q$
($i$ is the light-quark index,
the so-called `` Goldstone index "),
\begin{eqnarray}
H=(H_1, H_2, H_3)=(Q \bar u, Q \bar d, Q \bar s)
\label{eq:e2}
\end{eqnarray}
$H_i$ is also a $4\times 4$ matrix with respect to the Dirac index:
\begin{eqnarray}
H_i=\frac{1+\not\!v}{2} (-P_i \gamma_5+P_{i \mu}^\ast \gamma^\mu),\;
{\rm and}\; 

\bar H_i=\gamma_0 H^+_i \gamma_0
=(P^{+}_i \gamma_5+P_{i\mu}^{\ast +} \gamma^\mu)
\frac{1+\not\!v}{2},
\label{eq:e3}
\end{eqnarray}
where $P_i$ nad $P_{i \mu}^\ast$ are the pseudoscalar and vector
meson fields with velocity $v$. Note that $Tr_G$ implies a trace
over the Goldstone indices. Other notations are 

\begin{eqnarray}
D^\mu H_i=\partial^\mu H_i - H_jV^\mu_{ji}\ , 

V^\mu=\frac{1}{2} (\xi^+ \partial^\mu \xi+\xi \partial \xi^+ )\ ,
A^\mu=\frac{i}{2} (\xi^+ \partial^\mu \xi-\xi \partial \xi^+ )\ ,
\label{eq:e6}
\end{eqnarray}
where Goldstone bosons are described by $\xi$, defined as
\begin{eqnarray}
\xi=exp(\frac{iM}{f})\;\;( f\simeq 0.132\gev ),\ M=\!
 \left( \begin{array}{ccc}
\pi^0/\sqrt2+\eta/\sqrt6& \pi^+& K^+\\
\pi^-&-\pi^0/\sqrt2+\eta/\sqrt6&K^0 \\
K^-& \bar K^0 & -2/\sqrt6\eta
    \end{array}  \right) .
\label{eq:e9}
\end{eqnarray}

The third to the last term in (\ref{eq:e1}) represent interactions
associated with single heavy quark baryons.
Single heavy quark baryons can be classified by their light quark
content. Thus, antitriplet baryon field $B_{\bar 3}$
has spin $1/2$ while sextet baryon field $B_6$ has spin either
$1/2$ or $3/2$. With respect to Goldstone index one has,
for example, for antitriplet charmed baryons,
\begin{eqnarray}
B^{(c)}_{\bar 3}\!\!=\!\!
\left( \begin{array}{ccc}
0& \Lambda_c^+&\Xi^+_c\\
-\Lambda_c^+& 0 & \Xi^0_c \\
-\Xi^+_c & -\Xi^0_c& 0
 \end{array}  \right)\; {\rm or }\ \;
T_{i}=\frac{1}{2} \epsilon_{ijk}\ B_{\bar 3 jk},\;{\rm i.e }\ \;
T^{(c)}_1=\Xi^0_c,\ T^{(c)}_2=-\Xi^+_c,\ T^{(c)}_3=\Lambda^+_c\ .
\label{eq:e10}
\end{eqnarray}
The matrices for sextet baryons can also be easily written down.
The spin $1/2$ and $3/2$ sextet baryon fields
with the same quark content can be combined together:
\begin{eqnarray}
B_6^\mu=\frac{\gamma_\mu+v_\mu}{\sqrt3} \gamma_5 B_6(1/2)+
B^\mu_6(3/2)
\label{eq:e12}
\end{eqnarray}
The covariant derivative $D^\mu$ acts on baryon fields as
\begin{eqnarray}
D^\mu B_{ {\bar 3}, 6}=\partial^\mu B_{ {\bar 3}, 6}
+V^\mu B_{ {\bar 3}, 6}+B_{ {\bar 3}, 6} V^{\mu T} .
\label{eq:e13}
\end{eqnarray}
Under chiral SU(3)$_L$$\times$SU(3)$_R$ transformation, one finds 

\begin{eqnarray}
&&H \rightarrow H U^+, \ 

{\bar H} \rightarrow U {\bar H} , \ \;
B_{ {\bar 3}, 6} \rightarrow U B_{ {\bar 3}, 6} U^T, \ 

{\bar B}_{ {\bar 3}, 6} \rightarrow U^{T +}
{\bar B}_{ {\bar 3}, 6} U^+  \nonumber\\
&&\xi \rightarrow L\xi U^+=U \xi R^+, \ \ \ \
\xi^+ \rightarrow U \xi^+ L^+=R \xi^+ U^+ \nonumber\\
&& D^\mu H \rightarrow (D_\mu H)  U^+ , \ \
D^\mu B_{ {\bar 3}, 6} \rightarrow U (D^\mu B_{ {\bar 3}, 6}) U^T,
\ \ \ A^\mu \rightarrow U A^\mu U^+
\label{eq:e14}
\end{eqnarray}

\begin{center}
{\bf 3. Chiral Effective Hamiltonian of Weak Nonleptonic
Decays of Bottom Hadrons}
\end{center}

We consider here the Cabibbo-favored decays of type $\Delta b=-1,
\Delta c=0, \Delta s=1$, i.e. decays of type
$ b \rightarrow c {\bar c} s$ in the free quark case.
In the heavy limit for both $b$ and $c$ quarks, the effective
Hamiltonian reads \cite{r9}:
\begin{eqnarray}
H_{eff}=\frac{G_F V_{bc} V_{cs} }{\sqrt2}
( C_1^{\prime\prime} Q_1^{\prime\prime}+
C_2^{\prime\prime} Q_2^{\prime\prime}),
\phantom{abcdefghijklmij}\nonumber\\
O_1^{\prime\prime}=({\bar s} \Gamma_\mu h^{(b)}_v)
({\bar h^{(c)} }_{v^\prime} \Gamma^\mu h^{(\bar c)}_{\bar v}),
\ \ 

O_2^{\prime\prime}=({\bar s_\beta} \Gamma_\mu h^{(b)}_{\alpha v})
({\bar h^{(c)}}_{\alpha v^\prime}
\Gamma^\mu h^{(\bar c)}_{\beta \bar v}),
\label{eq:e15}
\end{eqnarray}
with $\Gamma_\mu=\gamma_\mu (1-\gamma_5)$.
In (\ref{eq:e15}), $\alpha$ and $\beta$ are color indices,
$C_1^{\prime\prime}$ and $C_2^{\prime\prime}$ are
Wilson coefficients, and $h^{(b)}_v$ indicates the heavy quark field
for $b$ quark.
It is obvious that $H_{eff}$ in (\ref{eq:e15}) transforms under
SU(3)$_L$$\times$SU(3)$_R$ as $(1_R, 1_L)\otimes(1_R, \bar 3_L)=
(1_R, \bar 3_L)$. Our task then is to construct the
lowest-order chiral effective Hamiltonian
on the hadron level which transforms as $(1_R, \bar 3_L)$.

\begin{center}
{\bf A. Bottom baryon decays}
\end{center}

We first study decays of type
$B^{(b)} \rightarrow B^{(c)}+\ds+\xi$
($\ds\!=\!D\ {\rm or}\ D^\ast$), where $B^{(b,c)}$
represent bottom and charmed baryons, and $\xi$ represents
Goldstone bosons. Let us consider the
Goldstone index for the moment.
According to (\ref{eq:e14}), we find that there are two possible
ways to form a
chiral effective Hamiltonian corresponding to (\ref{eq:e15}) which
transforms as $(1_R, \bar 3_L)$:
\begin{eqnarray}
H_{eff}\Rightarrow {\cal H}={\cal H}_1+{\cal H}_2,
\phantom{abcdefghijklmno}\nonumber\\
{\cal H}_1= Tr_G ( \bar B^{(c)} B^{(b)}) (H^{(c)}\xi^+)_i\ ,
\ \ \ \;
{\cal H}_2= ( H^{(c)} B^{(b)} {\bar B}^{(c)} \xi^+ )_i\ ,
\label{eq:e16}
\end{eqnarray}
where $i=3$ corresponds to the $s$ quark in (\ref{eq:e15}).
The baryon fields $B^{(c)}$ and $B^{(b)}$ could be either sextet
or antitriplet baryons.

For antitriple to antitriplet baryon decays
$B^{(b)}_{\bar 3} \rightarrow B^{(c)}_{\bar 3}+\ds+\xi$,
\begin{eqnarray}
&&{\cal H}_1=
Tr_G ( \bar B^{(c)}_{\bar 3} B^{(b)}_{\bar 3} ) (H^{(c)} \xi^+)_3
=-2\  {\bar T^{(c)} }_i H^{(c)}_j \xi^+_{j3} \ T^{(b)}_i, \nonumber\\
&&{\cal H}_2=
(H^{(c)} B^{(b)}_{\bar 3} \bar B^{(c)}_{\bar 3} \xi^+ )_3
= {\bar T^{(c)} }_j H^{(c)}_j \xi^+_{i3} \ T^{(b)}_i
 -{\bar T^{(c)} }_i H^{(c)}_j \xi^+_{j3} \ T^{(b)}_i\ .
\label{eq:e17}
\end{eqnarray}
Incorporating the heavy quark spin symmetry
we can write down the complete chiral effective Hamiltonian:
\begin{eqnarray}
&&{\cal H}_1\!\!
={\bar T^{(c)} }_i\ \Gamma_\mu\  H^{(c)}_j\  (A_1+B_1 {\not\! v})
\Gamma^\mu \ T^{(b)}_i \xi^+_{j3}\ ,\nonumber\\
&&{\cal H}_2=\!\!
{\bar T^{(c)} }_j \ \Gamma_\mu\
H^{(c)}_j\  (A_2+B_2 {\not\! v}) \ \Gamma^\mu\ T^{(b)}_i \xi^+_{i3}\  
, 

\label{eq:e24}
\end{eqnarray}
where $v$ is the velocity of the bottom baryon.
The heavy quark spin symmetry reduces the number of independent
amplitudes. Other types of decays such as 

$B^{(b)}_{\bar 3} \rightarrow B^{(c)}_{6}+ \ds+\xi$,
$B^{(b)}_{6} \rightarrow B^{(c)}_{\bar 3}+ \ds+\xi$
and  $B^{(b)}_{6} \rightarrow B^{(c)}_{6}+ \ds+\xi$,
can be similarly studied.

One can also study $B$-meson decays to charmed baryon and  
anti-charmed
baryon pair and Goldstone bosons using the same method \cite{rxk}.

\begin{center}
{\bf B. $B$-meson decays to two charmed mesons and Goldstone bosons}
\end{center}

For $B$-meson decays
$B \rightarrow \ds \ds_s  \xi$
produced by (\ref{eq:e15}) we again find, paying attention only to
Goldstone indices, there are two terms which
transform as $(1_R, \bar 3_L)$ under $SU(3)_L\times SU(3)_R$:
\begin{eqnarray}
{\cal H}={\cal H}_1+{\cal  
H}_2,\phantom{aaaaabcdefghijklmn}\nonumber\\
{\cal H}_1=
( H^{(b)} {\bar H}^{(c)} ) (H^{(c)} \xi^+)_3, \;\;\;
{\cal H}_2=
( H^{(c)} {\bar H}^{(c)} ) (H^{(b)} \xi^+)_3.
\label{eq:e35}
\end{eqnarray}
In \cite{r13}, it was found that heavy quark spin symmetry can not
reduce the number of independent amplitudes of decays
$B \rightarrow \ds \ds_s$. Thus, it is not useful
to write the chiral effective Hamiltonian for
$B \rightarrow  D^{(\ast)} D^{(\ast)}_s \xi$ in the compact form
as we did in (\ref{eq:e24}) for the baryon decays.

\begin{center}
{\bf 4. Decays $B^- \rightarrow  D^{+} D^{-}_s \pi^-$ and
$B^- \rightarrow  D^{\ast +} D^{-}_s \pi^-$ as examples }
\end{center}

Here we will calculate the decay distributions of
$B^- \rightarrow  D^{+} D^{-}_s \pi^-$ and
$B^- \rightarrow  D^{\ast +} D^{-}_s \pi^-$, as examples of
the application of chiral symmetry to many-body
nonleptonic decays of heavy hadrons. 

We will only consider the lowest-order contributions,
the commutator term and the pole term.
For $B^- \rightarrow  D^{+} D^{-}_s \pi^-$ and
$B^- \rightarrow  D^{\ast +} D^{-}_s \pi^-$,
the commutator term vanishes.

Let us consider $B^- \rightarrow  D^{+} D^{-}_s \pi^-$ as an
explicit example. There are two pole diagrams with
${\bar B}^{\ast 0}$ and
$D^{\ast 0}$ poles. We first define
\begin{eqnarray}
&&\langle D^+ D^-_s \mid {\cal H} \mid {\bar B}^{\ast 0} \rangle
=e^B_\mu {\cal M}^\mu_{D^+ D^-_s}\;,\;
\langle D^{\ast 0} D^-_s \mid {\cal H} \mid B^{-} \rangle
=e^{D \ast}_\mu {\cal N}^\mu_{D^{\ast 0} D^-_s}\;,
\label{eq:e38}
\end{eqnarray}
where $e^B_\mu$ and $e^D_\mu$ are polarization vectors of
${\bar B}^{\ast 0}$ and $D^{\ast 0}$. Then the two pole terms are
given by
\begin{eqnarray}
&&\langle D^+ D^-_s \pi^-\mid {\cal H} \mid B^-
\rangle_{ {\bar B}^{\ast 0}-{\textstyle pole} }
=i(\frac{g_1}{f} )
\frac{(p_\pi\cdot v)v_\mu-p_{\pi\mu} }{v\cdot p_\pi+\Delta_b}
{\cal M}^\mu_{D^+ D^-_s}, \nonumber\\
&&\langle D^+ D^-_s \pi^-\mid {\cal H} \mid B^-
\rangle_{ D^{\ast 0}-{\textstyle pole} }
=-i (\frac{g_1}{f} )
\frac{(p_\pi\cdot v^\prime)v^\prime_\mu-p_{\pi\mu} }
{v^\prime\cdot p_\pi - \Delta_c}
{\cal N}^\mu_{D^{\ast 0} D^-_s}.
\label{eq:e39}
\end{eqnarray}
where $\Delta_b=m_{B^\ast}-m_B$ and $\Delta_c=m_{D^\ast}-m_D$.

In our calculation we work in factorization approximation
and use heavy quark symmetry relations to
determine the amplitudes in (\ref{eq:e38}).
Thus, one can write ${\cal N}^\mu_{D^{\ast 0} D^-_s}$ and 

${\cal M}^\mu_{D^+ D^-_s}$ of (\ref{eq:e38}) as
\begin{eqnarray}
&&{\cal M}^\mu_{D^+ D^-_s}=i M_1\ p^\mu_{D^+}, \ \ \ \  

{\cal N}^\mu_{D^{\ast 0} D^-_s}=i M_2\ p^\mu_B \ .
\label{eq:e41}
\end{eqnarray}
One also has the heavy quark symmetry relation \cite{r13}
\begin{eqnarray}
\langle D^+ D^{\ast -}_s | H_{eff} | {\bar B}^{0} \rangle=
\langle D^+ D^-_s | H_{eff} | {\bar B}^{\ast 0} \rangle
\label{eq:e43}
\end{eqnarray}
With these relations and experimental data on
$B^-\rightarrow D^{\ast 0} D^-_s$ and  

${\bar B}^{0}\rightarrow D^+ D^{\ast -}_s$
(i.e we use data listed in \cite{r14})
we can determine the two-body decay amplitudes we need.
For $B^- \rightarrow  D^{\ast +} D^{-}_s \pi^-$
a similar calculation can be carried out.

We can now calculate the decay distributions of
$B^- \rightarrow  D^{+} D^{-}_s \pi^-$ and
$B^- \rightarrow  D^{\ast +} D^{-}_s \pi^-$. The branching ratio
distribution over $p_{12}=p^{(\ast)}_D+p_{D_s}$ and
$p_{13}=p^{(\ast)}_D+p_\pi$, for example, is given by
\begin{eqnarray}
dBr=\frac{\tau_B}{32 m_B^3 (2 \pi)^3} dp_{12}^2\ dp_{13}^2\!\!
\sum_{polarization}\!\!g_1^2 \ |{\hat A}|^2 \ \equiv g_1^2\ d{\hat  
Br}\ ,
\label{eq:e48}
\end{eqnarray}
where ${\hat A}$ is the decay amplitude with
$g_1$ being factored out, since there is only an upper limit for
$g_1$ coming from $D^{\ast}\rightarrow D\pi$ \cite{r1}. 

Chiral symmetry can obviously only be applied in
region where the pion is soft:
$p_{12}^2\simeq m_B^2, \ p_{13}^2\simeq m^2_D$.
In our calculation, we have taken into account
the width of $D^{\ast 0}$, i.e.
we have made the following replacement in
(\ref{eq:e39}):
$\frac{1}{\textstyle v^\prime\cdot p_\pi-\Delta_c} \rightarrow
 \frac{1}{\textstyle v^\prime\cdot
p_\pi-\Delta_c+i(\Gamma_{D^{\ast 0}}/2)}$.
We use $\Gamma_{D^{\ast 0}}=0.08$MeV, $\Delta_b=0.05$GeV and
$\Delta_b=0.14$GeV. The distributions $d {\hat Br}/d p_{12}^2$
and $d {\hat Br}/d p_{13}^2$ of
$B^- \rightarrow  D^{+} D^{-}_s \pi^-$ and
$B^- \rightarrow  D^{\ast +} D^{-}_s \pi^-$ are shown in
Figs.1-4. Though the results are
{\em only valid in the soft-pion region}, i.e.
near $p_{12}^2\simeq m_B^2$ and $p_{13}^2\simeq m^2_D$, we show the
distributions in the whole kinematic region, hoping to give an
order-of-magnitude estimate. Also as a rough estimate, we get
$Br(B^- \rightarrow  D^{+} D^{-}_s \pi^-)\approx 0.1 g^2_1\ \%$ and
$Br(B^- \rightarrow  D^{\ast +} D^{-}_s \pi^-)\approx 3 g^2_1\ \%$
from the above calculation.

\begin{picture}(0,200)(0,-130)
\includegraphics{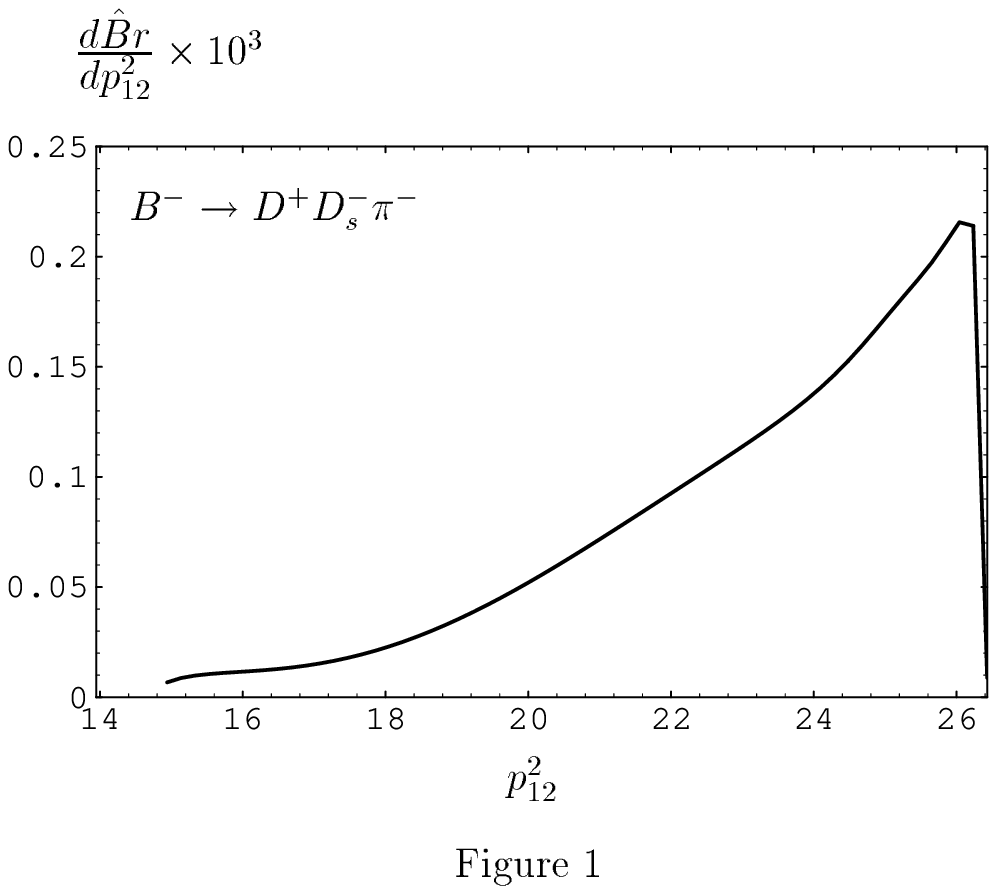}
\includegraphics{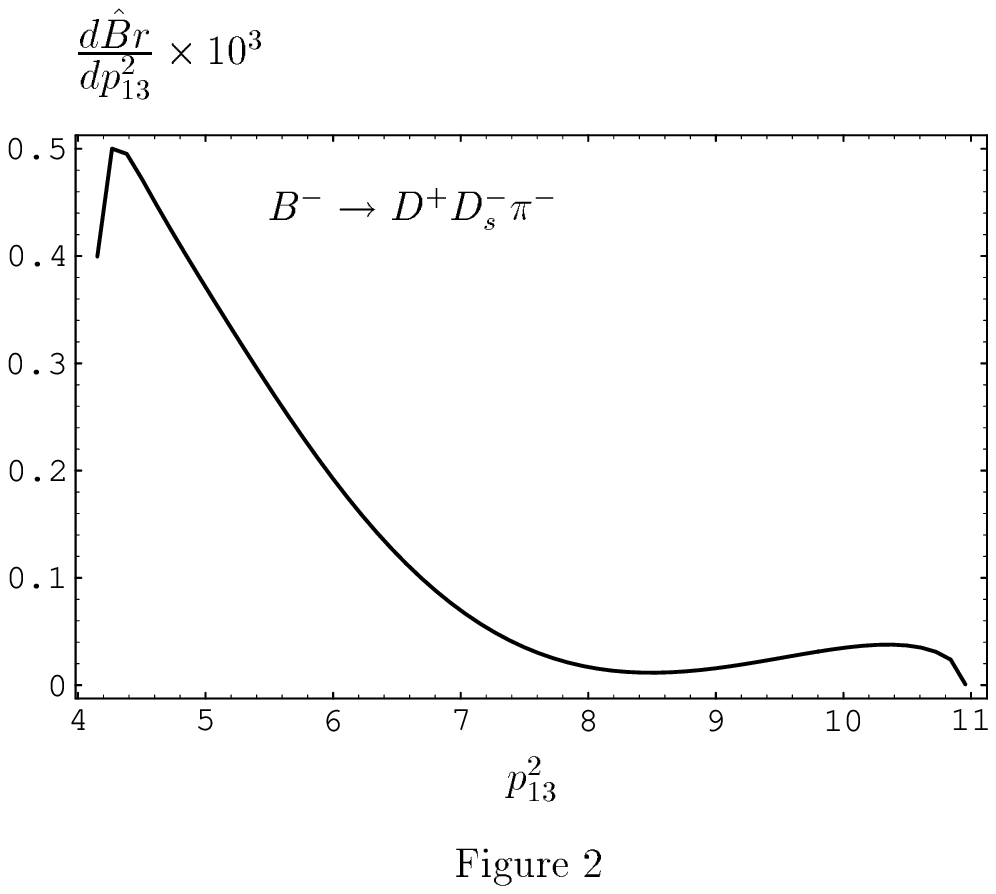}
\end{picture}
\pagebreak

\begin{picture}(0,250)(0,-170)
\includegraphics{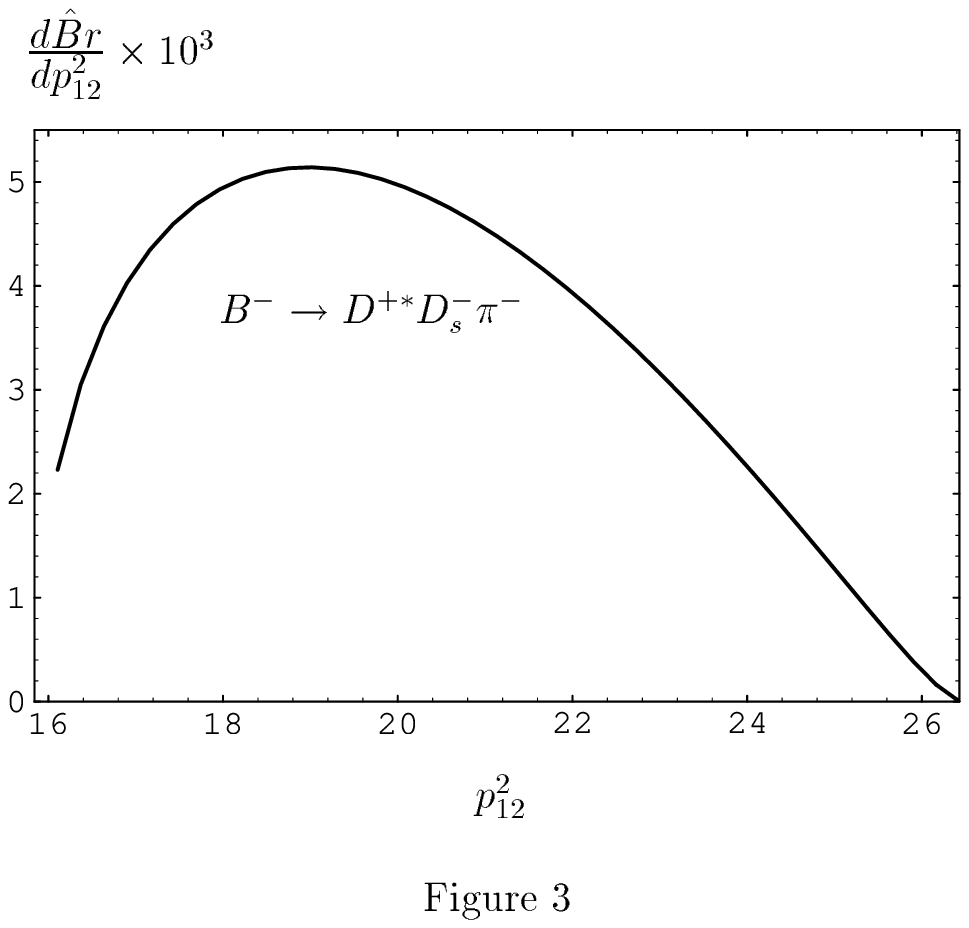}
\includegraphics{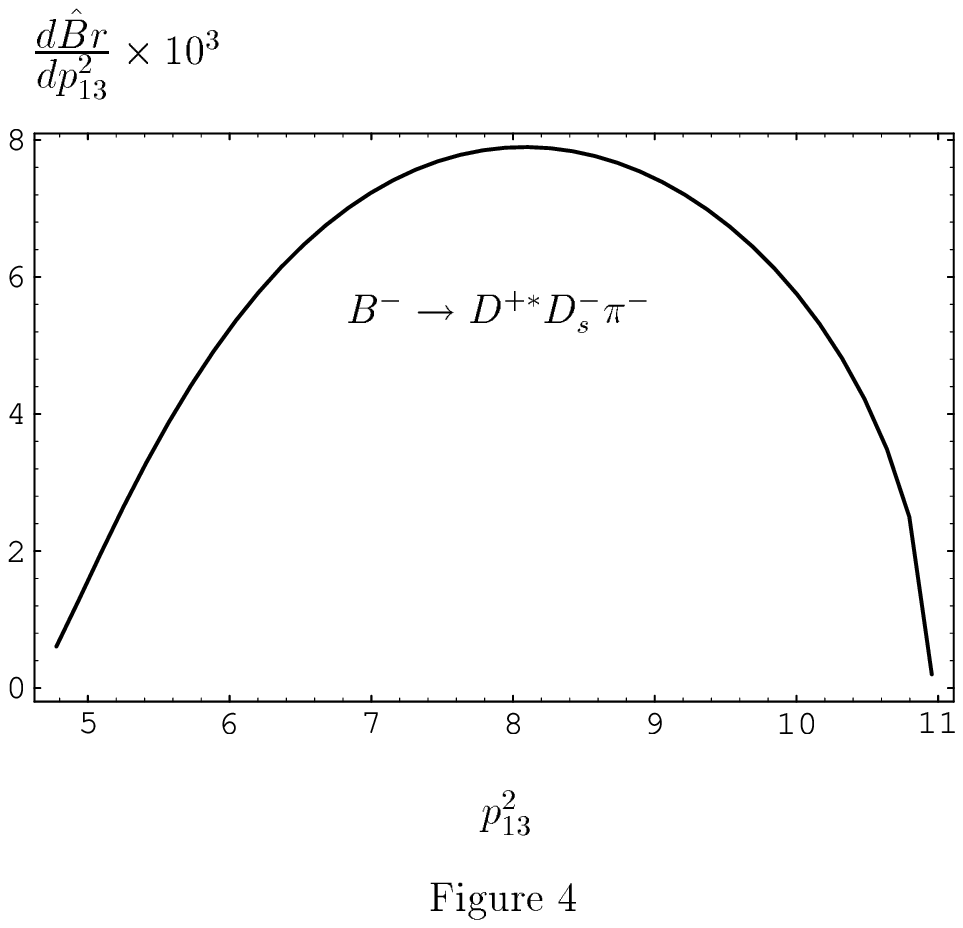}
\end{picture}

Our method can be applied to numerous other processes \cite{r15},
such as heavy to light decays $B^0 \ra P {\bar \Lambda} \pi$,
$B \ra D \rho \pi$, $\Lambda^+_c \ra P K^{\ast-} \pi^+$, and so on.
There are also some decays where the Goldstone bosons are soft
in the whole kinematic region and therefore chiral symmetry
can be applied to calculate
the full decay distribution and the decay rate, such as the measured
$D \ra a_1 K$ and $D^+ \ra K^+ K^+ K^-$.

Finally we want to emphasize that we have assumed the validity
of heavy quark symmetry and chiral symmetry.
It is certainly necessary, as the next
step, to investigate the symmetry-breaking effects. Also, the
contributions from pole terms of excited heavy mesons and baryons
may be important.

\vskip 0.1cm

This research was partly supported by a grant to A. N. Kamal from the
Natural Sciences and Engineering Research Council of Canada.

\end{document}